\documentclass[a4paper,11pt]{article}
\usepackage[utf8]{inputenc}
\usepackage[T1]{fontenc}
\usepackage[english]{babel}
\usepackage[hmargin={32mm,32mm},vmargin={32mm,35mm}]{geometry}
\usepackage{amsmath}
\usepackage{amsfonts}
\usepackage{amssymb}
\usepackage[mathscr]{euscript}
\usepackage{enumerate}
\usepackage{graphicx}
\usepackage{color}
\usepackage[hidelinks]{hyperref}
\usepackage{caption}
\usepackage{bookmark}
\usepackage{csquotes}
\usepackage[
	bibstyle=phys,
	biblabel=brackets,
	citestyle=numeric-comp,
	sorting=none,
	doi=false,
	eprint=true,
	pageranges=false,
	maxbibnames=10,
	backend=biber
]{biblatex}

\DeclareFieldFormat[article,inproceedings,inbook]{title}{\mkbibitalic{#1\isdot}}
\AtEveryBibitem{
	\ifentrytype{book}{
		\clearfield{series}
		\clearfield{volume}
	}{}
	\ifentrytype{inbook}{
		\clearfield{pages}
	}{}
}

\newcommand{\bra}[1]{\langle #1 |}
\newcommand{\ket}[1]{|#1\rangle}

\newcommand{\updown}[2]{^{#1}_{\phantom{#1}#2}}

\newcommand{\D}{{\mathscr{D}}}
\newcommand{\Rt}{\,{}^{(3)}\!R}

\numberwithin{equation}{section}

\begin{document}

\begin{center}

\Large
\textbf{Effective dynamics of a homogeneous and isotropic universe with quantum curvature}

\vspace{16pt}

\large
Ilkka Mäkinen

\normalsize

\vspace{12pt}

National Centre for Nuclear Research \\
Pasteura 7, 02-093 Warsaw, Poland

\vspace{8pt}

ilkka.makinen@ncbj.gov.pl

\end{center}

\renewcommand{\abstractname}{\vspace{-\baselineskip}}

\begin{abstract}
	\noindent We consider the effective dynamics of a new, tentative model of a homogeneous and isotropic universe in loop quantum cosmology. The new model consists of modifying the standard Hamiltonian of loop quantum cosmology by adding a Lorentzian term corresponding to the scalar curvature of the spatial manifold. The expression of the new Lorentzian term is motivated by a heuristic argument based on the form of an operator representing the scalar curvature in the one-vertex model of quantum-reduced loop gravity. The effective dynamics of the new model is not identical to standard LQC but all the key features of the dynamics are qualitatively reproduced. The classical cosmological singularity is resolved by a ``quantum bounce'' and the effective dynamics trajectory as a function of time is symmetric around the bounce; however, in comparison with the standard LQC scenario the bounce takes place at a significantly lower value of the volume.
\end{abstract}

\section{Introduction}

Among the attempts to apply methods derived from loop quantum gravity to the description of concrete physical situations, loop quantum cosmology (LQC) is arguably distinguished as the most well-developed and widely studied one (see e.g.~\cite{Ashtekar:2011ni, Agullo:2016tjh, Li:2023dwy} and references therein). In loop quantum cosmology, the phase space of a cosmological (e.g.~homogeneous and isotropic) universe is quantized using techniques inspired by the loop quantization of the full phase space of general relativity in the Ashtekar formulation. A key result of loop quantum cosmology is the resolution of the singularity present in the classical dynamics of a cosmological universe. Instead of a singularity, the quantum dynamics exhibits a ``bounce'' at a finite value of the volume of the universe, with a region of non-classical evolution around the bounce interpolating between an expanding classical trajectory in the far future and a contracting classical trajectory in the far past. The resolution of the singularity is well established both at the level of the full quantum dynamics, as well as the so-called effective dynamics \cite{Ashtekar:2006uz, Ashtekar:2006wn, Taveras:2008ke, Ding:2008tq}, which in this context refers to dynamics generated on the classical phase space by a semiclassical effective Hamiltonian, which is expected to capture at least some key features of the true quantum dynamics.

In this article we present a brief, preliminary look at a tentative new model of loop quantum cosmology, focusing on the case of a homogeneous and isotropic universe. The new model differs from standard loop quantum cosmology in that the Hamiltonian includes a non-trivial Lorentzian term corresponding to the scalar curvature of the spatial manifold. This proposal is nevertheless not inconsistent with the curvature of a spatially homogeneous universe being classically zero, as the new Lorentzian term approaches zero in the limit of vanishing ``polymerization parameter'' $\mu$. One can therefore argue that the new term describes quantum fluctuations of the curvature on top of its (zero) classical value. This approach represents a clear departure from standard loop quantum cosmology, where the Hamiltonian consists of the Euclidean part only, i.e.~one essentially assumes that the classically vanishing curvature is represented in the quantum theory by an identically zero operator.

The form of the new Lorentzian term is motivated by a heuristic argument based on the Hamiltonian constraint operator in the so-called one-vertex model of quantum-reduced loop gravity \cite{Makinen:2024rbg, Makinen:2026rof, Makinen:2026wwp}. The Euclidean part of the Hamiltonian in the one-vertex model is formally quite similar to the Hamiltonian used in standard loop quantum cosmology. By applying this formal analogy to the Lorentzian part of the Hamiltonian, which is given in the one-vertex model by an operator corresponding to the three-dimensional Ricci scalar \cite{Lewandowski:2021iun, Lewandowski:2022xox}, one can propose a plausible conjecture for an expression representing the scalar curvature in the setting of homogeneous and isotropic loop quantum cosmology \cite{Makinen:2024rbg}. However, it should be emphasized that, at least for the time being, this conjecture is of a purely heuristic nature; it should not currently be thought of as being the outcome of any concrete, systematic derivation from the formalism of full loop quantum gravity or loop quantum cosmology.

In the present work we will consider the new model in terms of its effective dynamics, utilizing a free scalar field as a relational clock for the dynamics of the gravitational degrees of freedom. The effective dynamics of the new model will be compared against standard LQC, as well a similar earlier proposal due to Dapor and Liegener \cite{Dapor:2017rwv, Assanioussi:2018hee, Assanioussi:2019iye}, where a Lorentzian term for the Hamiltonian in loop quantum cosmology is derived from Thiemann's regularization \cite{Thiemann:1996aw} of the Lorentzian part of the Hamiltonian in loop quantum gravity. We find that, as in standard LQC, the classical singularity is resolved and replaced with a bounce under the dynamics of the new model. The trajectory given by the new model is symmetric with respect to the bounce as a function of time, and agrees with the classical dynamics in the far future and far past of the bounce. The most significant difference with respect to the standard LQC dynamics seems to be that the bounce occurs at a distinctly lower value of the volume in the new model.

\section{The one-vertex model of quantum-reduced loop gravity}

The motivation for the cosmological model considered in this article comes from the form of the Hamiltonian constraint operator in the one-vertex model of quantum-reduced loop gravity \cite{Makinen:2024rbg, Makinen:2026rof, Makinen:2026wwp}. The one-vertex model is obtained from the general formalism of quantum-reduced loop gravity \cite{Alesci:2012md, Alesci:2013xd, Alesci:2013xya, Alesci:2016gub, Makinen:2020rda, Makinen:2023shj} by considering states defined on a cubic graph consisting of a single six-valent vertex. One assumes that the graph is embedded in a spatial manifold which is topologically a three-torus; thus, the graph is formed by three mutually orthogonal edges which close on themselves, such that the single vertex is both the beginning and ending point of each edge. An orthonormal basis on the Hilbert space of the one-vertex model is given by (generalized) spin network functions -- typically referred to as ``reduced spin network states'' in quantum-reduced loop gravity -- of the form
\begin{equation}
	\ket{j_xj_yj_z} = \D^{(j_x)}_{j_xj_x}(h_{e_x})_x \D^{(j_y)}_{j_yj_y}(h_{e_y})_y \D^{(j_z)}_{j_zj_z}(h_{e_z})_z.
	\label{eq:basis}
\end{equation}
Here the notation
\begin{equation}
	\D^{(j)}_{mn}(g)_i = \sqrt{2j+1}\, {}_i\bra{jm}D^{(j)}(g)\ket{jn}_i
	\label{}
\end{equation}
denotes the normalized matrix elements of the $SU(2)$ representation matrices $D^{(j)}(g)$ with respect to the basis $\ket{jm}_i$ ($i = x, y, z$), which diagonalizes the angular momentum operators $\hat J^2$ and $\hat J_i$.

The Hamiltonian constraint operator in the one-vertex model takes the form
\begin{equation}
	\hat H = \frac{1}{\beta^2}\hat H_E + \frac{1 + \beta^2}{\beta^2}\hat H_L
	\label{eq:H}
\end{equation}
where $\beta$ is the Barbero--Immirzi parameter and, up to questions of factor ordering, the Euclidean and Lorentzian parts of the operator are given by \cite{Makinen:2024rbg, Makinen:2026rof} 
\begin{equation}
	\hat H_E = -\sqrt{\frac{\hat p_x\hat p_y}{\hat p_z}}\hat s^{(1)}_x\hat s^{(1)}_y - \sqrt{\frac{\hat p_x\hat p_z}{\hat p_y}}\hat s^{(1)}_x\hat s^{(1)}_z - \sqrt{\frac{\hat p_y\hat p_z}{\hat p_x}}\hat s^{(1)}_y\hat s^{(1)}_z
	\label{eq:H_E}
\end{equation}
and
\begin{equation}
	\hat H_L = -16\Biggl[\frac{\hat p_x^{3/2}}{\sqrt{\hat p_y\hat p_z}}\bigl(\hat s^{(1/2)}_x\bigr)^4 + \frac{\hat p_y^{3/2}}{\sqrt{\hat p_x\hat p_z}}\bigl(\hat s^{(1/2)}_y\bigr)^4 + \frac{\hat p_z^{3/2}}{\sqrt{\hat p_x\hat p_y}}\bigl(\hat s^{(1/2)}_z\bigr)^4\Biggr].
	\label{eq:H_L}
\end{equation}
The elementary operators entering the expressions \eqref{eq:H_E} and \eqref{eq:H_L} are defined as follows. The reduced flux operators $\hat p_i$ are diagonal on the basis \eqref{eq:basis}:
\begin{equation}
	\hat p_i\ket{j_xj_yj_z} = j_i\ket{j_xj_yj_z}.
	\label{}
\end{equation}
The inverse factors of the flux operators in Eqs.~\eqref{eq:H_E} and \eqref{eq:H_L} emerge from the so-called Tikhonov regularization of the inverse volume operator; hence their action on the basis states \eqref{eq:basis} is defined by the prescription
\begin{equation}
	\frac{1}{\hat p_i}\ket{j_xj_yj_z} = \begin{cases}
		\dfrac{1}{j_i}\ket{j_xj_yj_z} & \text{if} \; j_i \neq 0 \\[2ex]
		0 & \text{if} \; j_i = 0
	\end{cases}
	\label{}
\end{equation}
Finally, the operator $\hat s_i^{(k)}$, which denotes a certain symmetric combination of the reduced holonomy operators, acts on the basis states as
\begin{equation}
	\hat s^{(k)}_x \ket{j_xj_yj_z} = \frac{1}{2i}\Bigl(\ket{j_x+k, j_y, j_z} - \ket{j_x-k, j_y, j_z}\Bigr)
	\label{eq:s_x}
\end{equation}
and similarly for $i=y$ and $i=z$.

Each operator \eqref{eq:H_E} and \eqref{eq:H_L} is derived from a corresponding operator defined in the framework of full loop quantum gravity, when these are applied on the reduced spin network functions \eqref{eq:basis}. The operator underlying the Euclidean operator $\hat H_E$ is a straightforward modification of a construction considered previously e.g.~in \cite{Assanioussi:2015gka, Assanioussi:2017tql, Yang:2015zda}, but in the setting of quantum-reduced loop gravity, the holonomy regularizing the curvature of the Ashtekar connection is taken according to a graph-preserving loop prescription on a fixed cubic graph. In turn, the Lorentzian operator $\hat H_L$ originates from an operator representing the scalar curvature of the spatial manifold \cite{Lewandowski:2021iun, Lewandowski:2022xox}. The operator defined by Eqs.~\eqref{eq:H}--\eqref{eq:H_L} therefore represents a quantization of the Hamiltonian constraint corresponding to the classical expression
\begin{equation}
	H = \frac{1}{\beta^2}\frac{\epsilon\updown{ij}{k}E^a_iE^b_jF_{ab}^k}{\sqrt{\det E}} + \frac{1 + \beta^2}{\beta^2}\sqrt{\det E}\Rt
	\label{}
\end{equation}
where the Lorentzian term is represented by the spatial Ricci scalar.

\section{A modified Hamiltonian for loop quantum cosmology}

One can observe that there is a definite formal similarity between the Euclidean operator \eqref{eq:H_E} and the Hamiltonian constraint in models of loop quantum cosmology, both in terms of the actual quantum operators and at the level of ``polymerized'' semiclassical expressions, such as
\begin{equation}
	H_E^{(\mu)} = -\frac{3}{\beta^2}\sqrt p\frac{\sin^2\mu c}{\mu^2}
	\label{eq:H_E^mu}
\end{equation}
representing the Hamiltonian constraint of a spatially homogeneous and isotropic universe. (The similarity is particularly direct in the case of Bianchi I models describing an anisotropic universe; however, in this article we will restrict our attention to the homogeneous and isotropic case for simplicity.) The expression \eqref{eq:H_E^mu} is to be understood as a modification of the classical Hamiltonian constraint of a homogeneous and isotropic universe, in which the Ashtekar connection and the densitized triad have the form
\begin{equation}
	A_a^i = c(t)\delta_a^i, \qquad E^a_i = p(t)\delta^a_i,
	\label{}
\end{equation}
and the Hamiltonian constraint is given by
\begin{equation}
	H_{\rm cl} = -\frac{3}{\beta^2}\sqrt p c^2.
	\label{eq:H_cl}
\end{equation}
Eq.~\eqref{eq:H_E^mu} can be obtained from Eq.~\eqref{eq:H_cl} by making the replacement $c \to \sin \mu c / \mu$, which is referred to as ``polymerization'' of the connection variable $c$. Here the physical role of the parameter $\mu$ is essentially to encode the fact that spatial geometry is fundamentally discrete in loop quantum gravity, with the classical expression \eqref{eq:H_cl} being recovered from its polymerized counterpart \eqref{eq:H_E^mu} in the limit $\mu \to 0$.

The polymerization of the Hamiltonian constraint renders it suitable for quantization, as no operator corresponding to the connection $c$ exists in loop quantum cosmology, while the holonomy $e^{i\mu c}$ is available as a well-defined operator. Hence the sine of the connection in Eq.~\eqref{eq:H_E^mu} is quantized as
\begin{equation}
	\widehat{\sin\mu c} = \frac{1}{2i}\Bigl(\widehat{e^{i\mu c}} - \widehat{e^{-i\mu c}}\Bigr),
	\label{}
\end{equation}
which is formally very similar to the operator \eqref{eq:s_x}, since the holonomy acts as a shift operator on the standard basis of the LQC Hilbert space. Conversely, we can heuristically imagine obtaining the expression \eqref{eq:H_E^mu} from the Euclidean operator \eqref{eq:H_E} by means of some procedure where the holonomy and flux operators become substituted with their classical values; in particular, the sine-like operators $\hat s_i^{(1)}$ are replaced with $\sin\mu c$.

Heuristic as it may be, this analogy between the Hamiltonian in loop quantum cosmology and the Euclidean Hamiltonian of the one-vertex model opens up the possibility of proposing a modified Hamiltonian constraint for loop quantum cosmology by way of extending the analogy to the Lorentzian part of the Hamiltonian \cite{Makinen:2024rbg}. This amounts to writing down a polymerized expression representing the Lorentzian part of the Hamiltonian, which is related to the Lorentzian operator \eqref{eq:H_L} in the same way as the polymerized expression \eqref{eq:H_E^mu} is related to the Euclidean operator \eqref{eq:H_E}. In this way we are led to the (tentative) proposal
\begin{equation}
	H_L^{(\mu)} = -48\frac{1 + \beta^2}{\beta^2}\sqrt p\frac{\sin^4(\mu c/2)}{\mu^2}
	\label{eq:H_L^mu}
\end{equation}
for the Lorentzian term. Recalling that the Lorentzian operator \eqref{eq:H_L} originates from an operator corresponding to the three-dimensional scalar curvature, the interpretation of the new proposal would be that in the resulting quantum theory the spatial curvature will be represented by a non-trivial operator, in contrast to the standard treatment in loop quantum cosmology, where the classically vanishing curvature of a homogeneous and isotropic universe is assumed to be represented by an identically zero quantum operator. The expression \eqref{eq:H_L^mu} nevertheless approaches zero in the limit $\mu\to 0$, so the new proposal is at least not immediately inconsistent with the fact that the spatial curvature of a homogeneous and isotropic universe is classically zero.

It is important to emphasize that, while the expression \eqref{eq:H_L^mu} is a potentially interesting conjecture, it is not supported by any solid, systematic derivation from the formalism of loop quantum gravity for the time being. The situation is significantly different for the Euclidean part of the Hamiltonian, for which it has been shown via concrete calculations that an expression of the form \eqref{eq:H_E^mu} can be recovered as the expectation value of the Euclidean Hamiltonian constraint with respect to a family of semiclassical states peaked on homogeneous and isotropic classical data \cite{Dapor:2017gdk, Zhang:2021qul}. That is,
\begin{equation}
	\bra{\psi_{c, p}}\hat H_E\ket{\psi_{c, p}} = H_E^\mu(c, p) + {\cal O}(\hbar),
	\label{}
\end{equation}
where $\hat H_E$ now denotes the Hamiltonian constraint of full loop quantum gravity, and not the one-vertex operator \eqref{eq:H_E}. In order to put the proposal \eqref{eq:H_L^mu} on a more solid footing, one could attempt to establish a similar result for the curvature operator of \cite{Lewandowski:2021iun}. In fact, we expect that the correct setting for reproducing the expression \eqref{eq:H_L^mu} with the factors of $\mu$ included would be a semiclassical state defined on a large cubic graph. The role of the one-vertex model in this respect is merely to provide a convenient means of anticipating what the result of a proper derivation would look like even before one has embarked on the full calculation, which is likely to be rather long and elaborate.

We should also point out that the new proposal encoded in Eq.~\eqref{eq:H_L^mu} is quite similar in spirit to an earlier model which was introduced by Dapor and Liegener in \cite{Dapor:2017rwv}, and various aspects of which have been subsequently studied in the literature (see e.g.~\cite{Assanioussi:2018hee, Assanioussi:2019iye, Agullo:2018wbf, deHaro:2018khb, Gomar:2020orw, Garcia-Quismondo:2020wna}); yet another similar construction was considered still earlier in \cite{Yang:2009fp}. The central idea underlying both the Dapor--Liegener model and the new proposal discussed in this article is to use the Lorentzian part of the Hamiltonian constraint in full loop quantum gravity as a motivation for introducing a Lorentzian term for the Hamiltonian of loop quantum cosmology. In the case of the Dapor--Liegener model, the authors took a Hamiltonian operator corresponding to the classical expression
\begin{equation}
	H = \frac{E^a_iE^b_j}{\sqrt{\det E}}\Bigl(\epsilon\updown{ij}{k}F_{ab}^k - (1 + \beta^2)\bigl(K_a^iK_b^j - K_a^jK_b^i\bigr)\Bigr)
	\label{}
\end{equation}
with the Lorentzian part written in terms of the extrinsic curvature, the quantization of the Lorentzian term being given by the well-known construction due to Thiemann \cite{Thiemann:1996aw}. Looking at the expectation values of the Hamiltonian in homogeneous and isotropic coherent states, they showed that the Lorentzian part gives a contribution proportional to $\sin^2(2\mu c) / \mu^2$. Using elementary trigonometric identities, one can then put the total Hamiltonian of the model in the form
\begin{equation}
	H^{(\mu)}_{\rm DL} = -\frac{3}{\beta^2}\sqrt p\frac{\sin^2\mu c}{\mu^2} + 3\frac{1 + \beta^2}{\beta^2}\sqrt p\frac{\sin^4\mu c}{\mu^2},
	\label{}
\end{equation}
which consists of a different Lorentzian ``correction'' term applied over the standard LQC Hamiltonian \eqref{eq:H_E^mu}.

\section{Effective dynamics}

In this article we will be satisfied with examining the effective dynamics of the new model defined by Eq.~\eqref{eq:H_L^mu} and comparing it against standard loop quantum cosmology and the Dapor--Liegener model; an analysis of the true quantum dynamics of the model is left as a question for future work. A detailed discussion of the subject would be outside the scope of this article, but roughly speaking, effective dynamics \cite{Ashtekar:2006uz, Ashtekar:2006wn, Taveras:2008ke, Ding:2008tq} in the context of loop quantum cosmology and related models refers to the dynamics generated on the classical phase space by a semiclassical ``effective Hamiltonian'', where the contribution of the gravitational degrees of freedom is represented by a polymerized expression such as \eqref{eq:H_E^mu}. The general expectation is that the effective dynamics will accurately describe at least some relevant aspects of the actual quantum dynamics of the model. (See e.g.~the review articles \cite{Ashtekar:2011ni, Agullo:2016tjh, Li:2023dwy} for a more comprehensive discussion.)

We consider a homogeneous and isotropic gravitational field coupled to a free massless scalar field $\phi$. The effective Hamiltonian of the system is then given by
\begin{equation}
	H_{\rm eff} = H_{\rm gr}(c, p) + \frac{\pi_\phi^2}{2p^{3/2}},
	\label{}
\end{equation}
where $H_{\rm gr}(c, p)$ denotes the effective Hamiltonian of the gravitational degrees of freedom, and $\pi_\phi$ is the canonical momentum of the scalar field. In units where $8\pi G = 1$, the homogeneous connection and triad satisfy the Poisson bracket
\begin{equation}
	\{c, p\} = \frac{\beta}{3}
	\label{}
\end{equation}
while the Poisson bracket for the scalar field variables is
\begin{equation}
	\{\phi, \pi_\phi\} = 1.
	\label{}
\end{equation}
For a phase space function $F(c, p)$ depending on the gravitational variables only, Hamilton's equation of motion reads
\begin{equation}
	\dot F(c, p) = \{F, H_{\rm eff}\} = \{F, H_{\rm gr}\} - \frac{\beta}{4}\frac{\pi_\phi^2}{p^{5/2}}\frac{\partial F}{\partial c},
	\label{eq:dotF}
\end{equation}
where the dot denotes derivative with respect to the coordinate time $t$. The equations of motion for the scalar field are
\begin{align}
	\dot\phi &= \{\phi, H_{\rm eff}\} = \frac{\pi_\phi}{p^{3/2}}, \\[1ex]
	\dot\pi_\phi &= \{\pi_\phi, H_{\rm eff}\} = 0;
	\label{}
\end{align}
thus, the momentum $\pi_\phi$ is a constant of motion. In the general-relativistic setting, it is natural to use the scalar field as a relational time variable for the dynamics of the gravitational field. In this approach, the equation of motion of the gravitational variables with respect to the scalar field time is
\begin{equation}
	\frac{d}{d\phi}F(c, p) = \frac{\dot F}{\dot\phi} = \frac{p^{3/2}}{\pi_\phi}\dot F
	\label{}
\end{equation}
with $\dot F$ given by Eq.~\eqref{eq:dotF}.

\subsection{Standard LQC}

For the effective dynamics corresponding to standard loop quantum cosmology, the gravitational effective Hamiltonian is given by Eq.~\eqref{eq:H_E^mu}, i.e.
\begin{equation}
	H_{\rm gr} = -\frac{3}{\beta^2}\sqrt p\frac{\sin^2\mu c}{\mu^2}.
	\label{}
\end{equation}
We then obtain the equations of motion for the gravitational variables as
\begin{align}
	\dot p &= \frac{2}{\beta}\sqrt p\frac{\sin 2\mu c}{2\mu}, \label{eq:dp-LQC} \\
	\dot c &= -\frac{1}{2\beta}\frac{1}{\sqrt p}\frac{\sin^2\mu c}{\mu^2} - \frac{\beta}{4}\frac{\pi_\phi^2}{p^{5/2}}. \label{eq:dc-LQC}
\end{align}
The total Hamiltonian constraint $H = H_{\rm gr} + \pi_\phi^2/2p^{3/2} = 0$ implies that there is a constant of motion:
\begin{equation}
	\biggl(p\frac{\sin\mu c}{\mu}\biggr)^2 = \frac{\beta^2}{6}\pi_\phi^2.
	\label{}
\end{equation}
Since $|\sin\mu c| \leq 1$, this shows that the value of the triad is bounded from below, the minimum value being
\begin{equation}
	p_{\rm min} = \frac{\beta\mu}{\sqrt 6}\pi_\phi.
	\label{}
\end{equation}
Hence the singularity present in the classical dynamics of a homogeneous and isotropic universe is resolved, and the effective dynamics trajectory instead features a ``bounce'' at a finite value of the volume $v = p^{3/2}$: 
\begin{equation}
	v_{\rm min} = \biggl(\frac{\beta\mu}{\sqrt 6}\pi_\phi\biggr)^{3/2}.
	\label{eq:v_min}
\end{equation}
A representative example of the standard LQC effective dynamics, obtained by numerically integrating the equations of motion, is shown in Figs.~\ref{fig:1}--\ref{fig:3} by the solid blue curve\footnote{
	The trajectory displayed in the plots corresponds to the value $\pi_\phi = 200$ of the scalar field momentum. The Barbero--Immirzi parameter is set to $\beta = 0.2375$, which is the standard value used in loop quantum cosmology \cite{Ashtekar:2011ni, Agullo:2016tjh, Li:2023dwy}. For the parameter $\mu$ we use the constant value $\mu = 1$, i.e.~we work in the $\mu_0$ scheme, although at the level of effective dynamics it would be straightforward to extend the analysis to the $\bar\mu$ scheme, where $\mu$ is taken to be a phase-space dependent function proportional to $1/\sqrt p$.
}. Aside from the presence of the bounce, the essential features of the dynamics are that the trajectory as a function of time is symmetric with respect to the bounce, and agrees with the classical trajectory (plotted as a black dashed curve) in the far future and far past of the bounce.

\subsection{The new model}

For the new model, where the expression \eqref{eq:H_L^mu} is added as a Lorentzian term to the standard LQC Hamiltonian \eqref{eq:H_E^mu}, the gravitational effective Hamiltonian is
\begin{equation}
	H_{\rm gr} = -\frac{3}{\beta^2}\sqrt p\biggl(\frac{\sin^2\mu c}{\mu^2} + 16(1 + \beta^2) \frac{\sin^4(\mu c/2)}{\mu^2}\biggr).
	\label{eq:H_new}
\end{equation}
Introducing the abbreviations
\begin{equation}
	\lambda = \frac{\mu}{2}, \qquad \alpha = 4(1 + \beta^2) - 1
	\label{}
\end{equation}
the Hamiltonian \eqref{eq:H_new} can be written as
\begin{equation}
	H_{\rm gr} = -\frac{3}{\beta^2}\sqrt p\frac{\sin^2\lambda c}{\lambda^2}\Bigl(1 + \alpha\sin^2\lambda c\Bigr),
	\label{}
\end{equation}
which yields the equations of motion
\begin{align}
	\dot p &= \frac{2}{\beta}\sqrt p\frac{\sin 2\lambda c}{2\lambda}\Bigl(1 + \alpha\bigl(1 - \cos 2\lambda c\bigr)\Bigr), \\
	\dot c &= -\frac{1}{2\beta}\frac{1}{\sqrt p}\frac{\sin^2\lambda c}{\lambda^2}\Bigl(1 + \alpha\sin^2\lambda c\Bigr) - \frac{\beta}{4}\frac{\pi_\phi^2}{p^{5/2}}.
	\label{}
\end{align}
The constraint $H_{\rm gr} + H_\phi = 0$ now shows that the quantity
\begin{equation}
	p^2\biggl(\frac{\sin^2\lambda c}{\lambda^2} + \alpha\frac{\sin^4\lambda c}{\lambda^2}\biggr) = \frac{\beta^2}{6}\pi_\phi^2
	\label{}
\end{equation}
is a constant of motion. Here the function within the parentheses on the left-hand side takes its maximum value $(1 + \alpha) / \lambda^2$ when $\sin^2 \lambda c = 1$. It follows that the value of the triad is bounded from below by
\begin{equation}
	p_{\rm min} = \frac{1}{\sqrt{1 + \alpha}}\frac{\beta\lambda}{\sqrt 6}\pi_\phi = \frac{1}{4\sqrt{1 + \beta^2}}\frac{\beta\mu}{\sqrt 6}\pi_\phi.
	\label{}
\end{equation}
Accordingly, the minimum value of the volume is
\begin{equation}
	v_{\rm min} = \frac{1}{8(1 + \beta^2)^{3/4}}v_{\rm min}^{(0)},
	\label{}
\end{equation}
where $v_{\rm min}^{(0)}$ denotes the value given by Eq.~\eqref{eq:v_min}. That is, under the dynamics of the new model the bounce occurs at a lower volume in relation to the standard LQC dynamics. For $\beta = 0.2375$ we have the numerical value $v_{\rm min} = 0.120v_{\rm min}^{(0)}$.

\begin{figure}[t]
	\centering
	\includegraphics[scale=0.5]{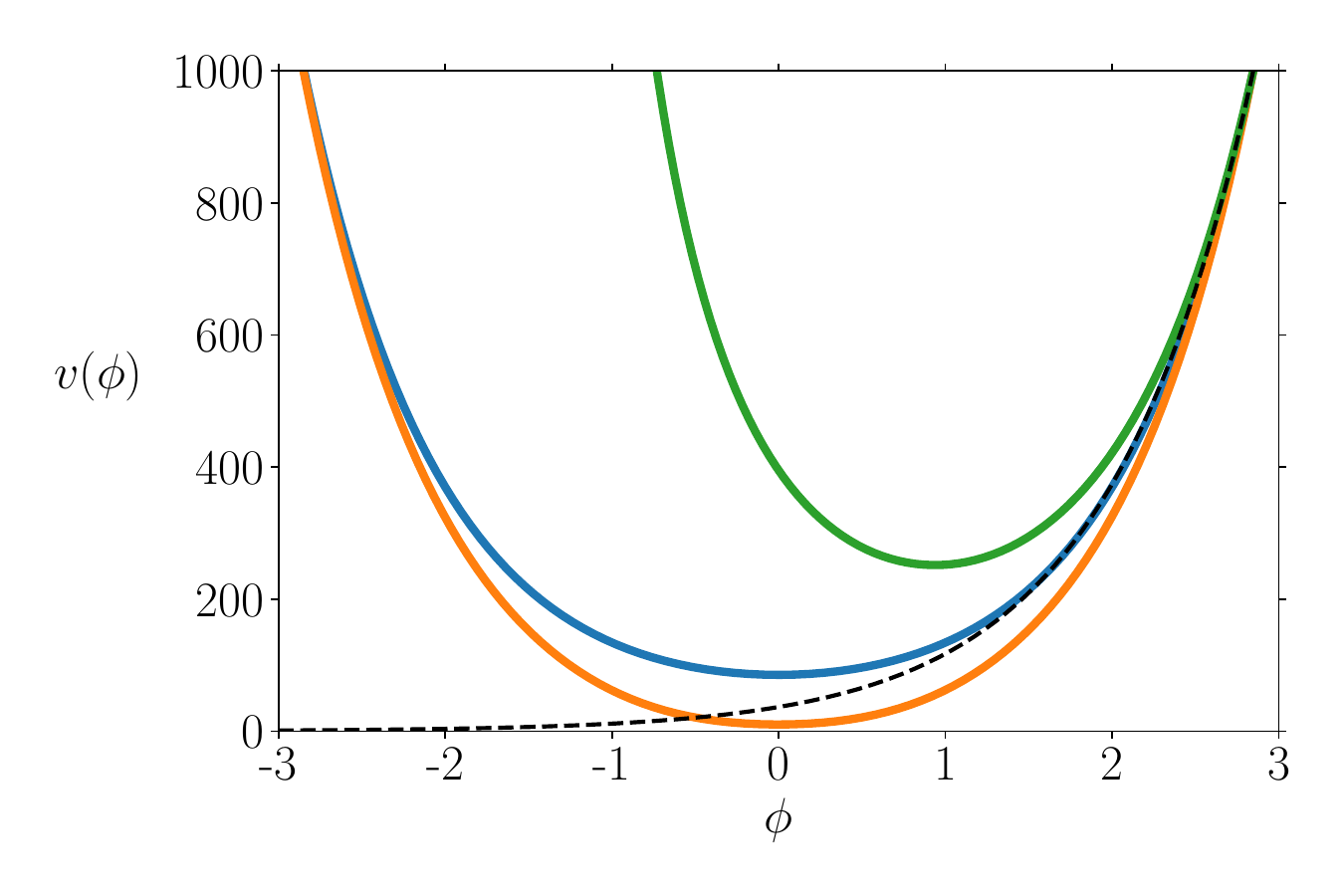}
	\caption{Effective dynamics of the volume $v = p^{3/2}$ as a function of the scalar field time $\phi$. Blue curve: Standard LQC. Orange curve: The new model. Green curve: The Dapor--Liegener model. The dashed black curve shows the classical trajectory, which terminates in a singularity at a finite coordinate time $t$.}
	\label{fig:1}
\end{figure}

The numerical solution of the equations of motion for the new model is shown by the orange curve in Figs.~\ref{fig:1}--\ref{fig:3}. Like the standard LQC trajectory, the trajectory derived from the new model is also symmetric with respect to the instant of the bounce, and converges to the classical trajectory in the far future and far past. As best seen from Fig.~\ref{fig:2}, the trajectory of the new model is slightly shifted relative to the standard LQC trajectory in the far past of the bounce. This indicates that the total duration of the ``quantum phase'', which interpolates between the classical trajectories in the past and future of the bounce, is slightly shorter in the new model in comparison with the standard LQC scenario.

\subsection{Dapor--Liegener model}

Finally we consider the Dapor--Liegener model, for which the effective gravitational Hamiltonian is given by
\begin{equation}
	H_{\rm gr} = -\frac{3}{\beta^2}\sqrt p\biggl(\frac{\sin^2\mu c}{\mu^2} - (1 + \beta^2)\frac{\sin^4\mu c}{\mu^2}\biggr).
	\label{}
\end{equation}
The corresponding equations of motion are
\begin{align}
	\dot p &= \frac{2}{\beta}\sqrt p\frac{\sin 2\mu c}{2\mu}\Bigl(1 - 2(1 + \beta^2)\sin^2\mu c\Bigr), \\
	\dot c &= -\frac{1}{2\beta}\frac{1}{\sqrt p}\frac{\sin^2\mu c}{\mu^2}\Bigl(1 - (1 + \beta^2)\sin^2\mu c\Bigr) - \frac{\beta}{4}\frac{\pi_\phi^2}{p^{5/2}}.
	\label{}
\end{align}
From the vanishing of the total Hamiltonian constraint we now deduce the constant of motion
\begin{equation}
	p^2\biggl(\frac{\sin^2\mu c}{\mu^2} - (1 + \beta^2)\frac{\sin^4\mu c}{\mu^2}\Bigr) = \frac{\beta^2}{6}\pi_\phi^2.
	\label{}
\end{equation}
To find the corresponding lower bound for the triad, we note that the maximum value of the function $f(x) = x - (1 + \beta^2)x^2$ is given by $1/4(1 + \beta^2)$. Thus the lower bound is
\begin{equation}
	p_{\rm min} = 2\sqrt{1 + \beta^2}\frac{\beta\mu}{\sqrt 6}\pi_\phi
	\label{}
\end{equation}
and hence the volume at the bounce reaches the value
\begin{equation}
	v_{\rm min} = 2\sqrt 2(1 + \beta^2)^{3/4}v_{\rm min}^{(0)},
	\label{}
\end{equation}
which is higher than the value $v_{\rm min}^{(0)}$ under the standard LQC dynamics. For $\beta = 0.2375$ we find $v_{\rm min} = 2.947v_{\rm min}^{(0)}$.

\begin{figure}[p]
	\centering
	\includegraphics[scale=0.5]{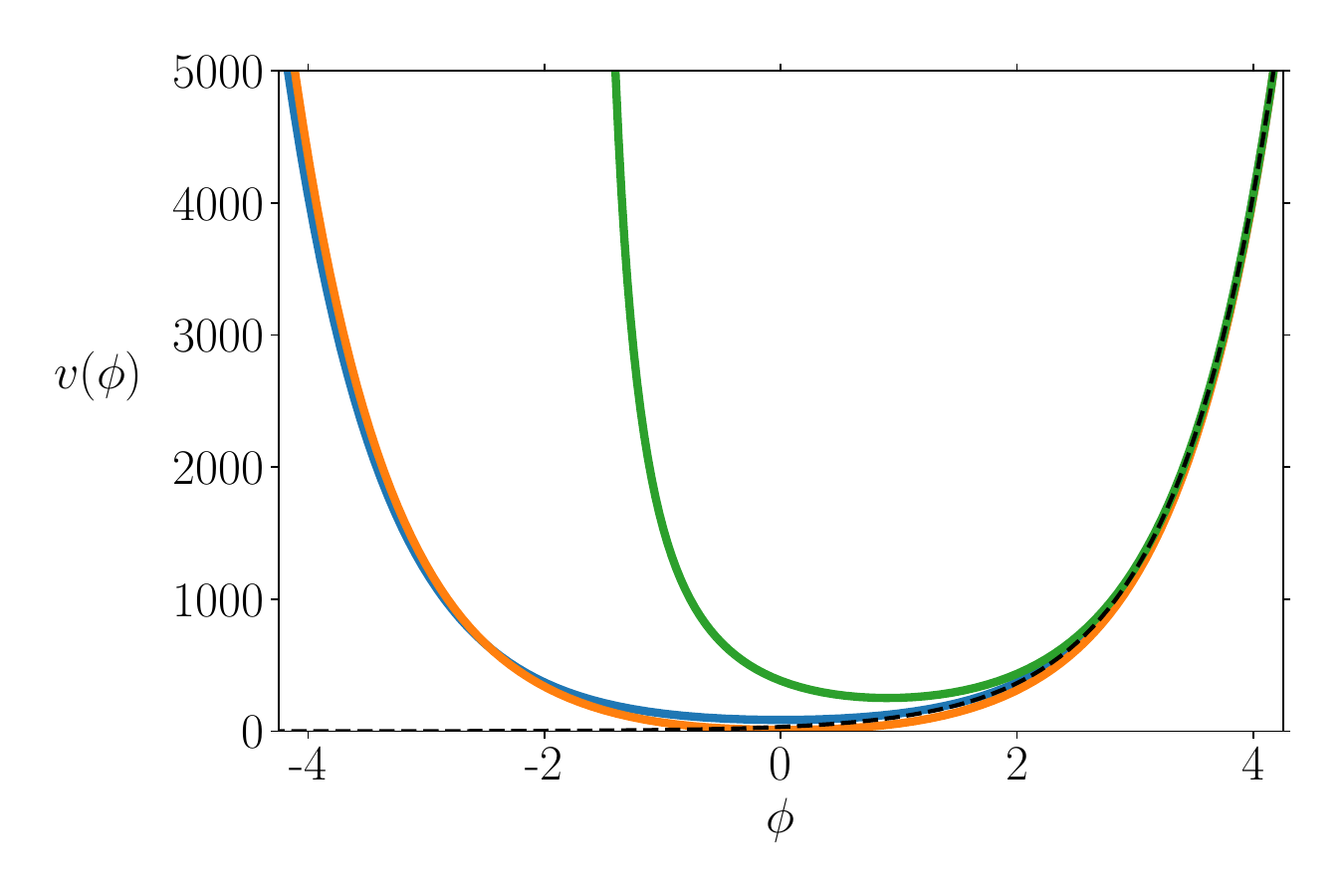}
	\caption{The effective dynamics of the volume over a longer time interval. All three trajectories agree with the classical dynamics in the far future of the bounce. The trajectories of standard LQC and the new model are symmetric with respect to the past and future of the bounce, and thus agree with a time-reversed classical trajectory also in the far past of the bounce. In contrast, the Dapor--Liegener trajectory is highly asymmetric, exhibiting a phase of exponential contraction in the past of the bounce. Note that the trajectory of the new model is slightly shifted with respect to the standard LQC trajectory in the far past, indicating that the duration of the non-classical phase around the bounce is slightly shorter in the case of the new model.}
	\label{fig:2}
\end{figure}

\begin{figure}[p]
	\centering
	\includegraphics[scale=0.5]{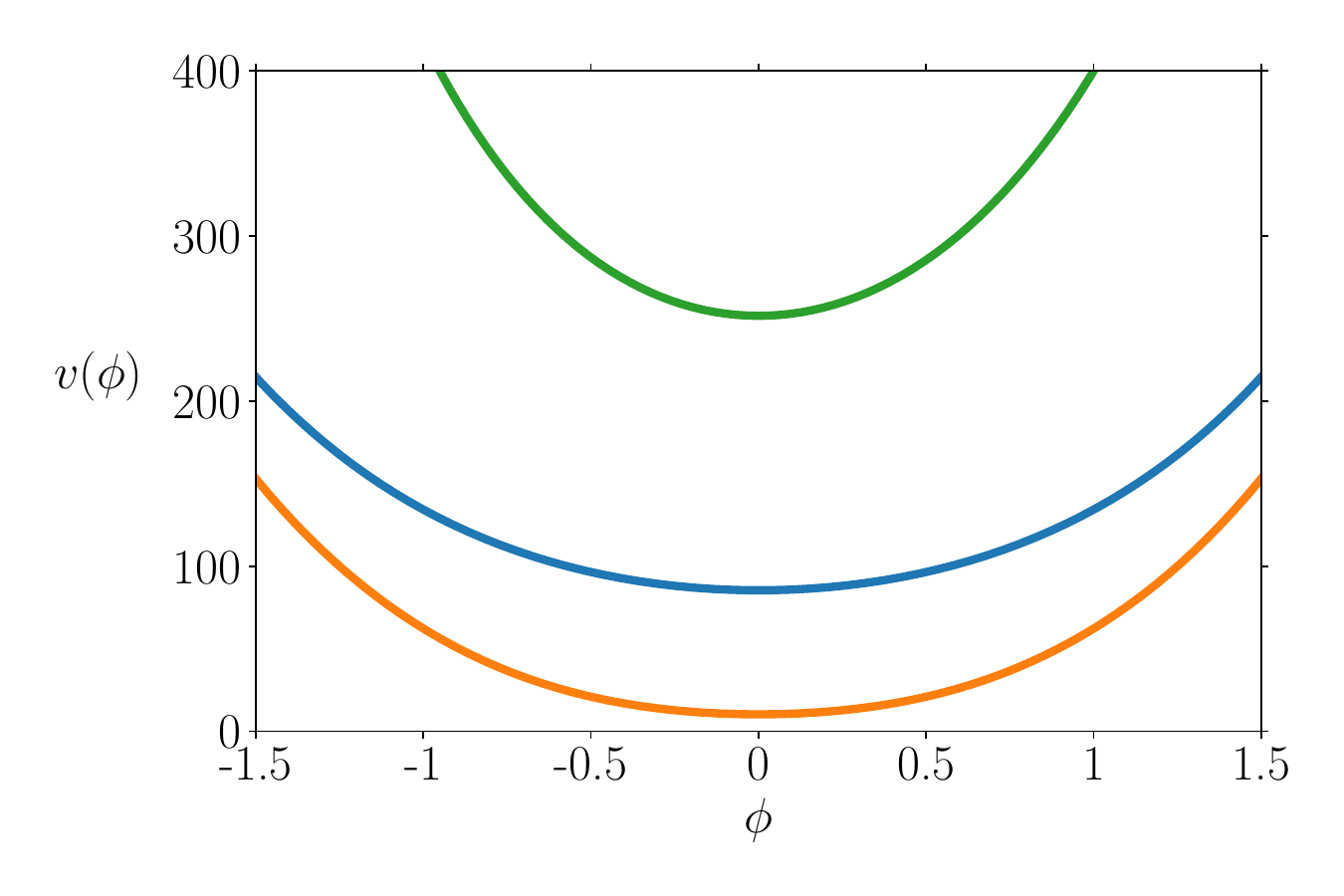}
	\caption{A closer view of the trajectories in the region near the bounce. In this plot the time axis for each trajectory has been shifted so that the bounce takes place at $\phi = 0$. Under the dynamics of the new model, the bounce occurs at a considerably lower value of the volume in comparison with the bounce in standard LQC. In the Dapor--Liegener model, the value of the volume at the bounce is higher than in standard LQC.}
	\label{fig:3}
\end{figure}

For the Dapor--Liegener model, the numerical solution of the equations of motion is plotted as the green curve in Figs.~\ref{fig:1}--\ref{fig:3}. Unlike the standard LQC trajectory and the solution obtained from the new model, the dynamics of the Dapor--Liegener model is not symmetric under time reversal around the bounce. For a solution which agrees with the classical trajectory in the far future, the dynamics in the past of the bounce exhibits a de Sitter-like phase characterized by an exponential contraction. Further analysis of the asymmetric dynamics of the Dapor--Liegener model can be found e.g.~in \cite{Agullo:2018wbf, deHaro:2018khb, Assanioussi:2019iye}.

\section{Conclusions}

In this article we examined a new, tentative model, originally proposed in \cite{Makinen:2024rbg}, describing a homogeneous and isotropic universe in the setting of loop quantum cosmology. The new model is obtained by amending the standard LQC Hamiltonian constraint with a Lorentzian term corresponding to the scalar curvature of the spatial manifold. This approach differs from the usual treatment of the Hamiltonian in loop quantum cosmology, which essentially amounts to assuming that the scalar curvature is represented by an identically vanishing operator in the quantum theory. The new Lorentzian term does approach zero in the limit of vanishing polymerization parameter $\mu$, so it presumably describes some kind of quantum fluctuations of the curvature, while not contradicting the requirement that the spatial curvature of a homogeneous and isotropic universe should reduce to zero in the classical limit.

The motivation for the new Lorentzian term stems from considering a certain formal analogy between the Hamiltonian constraint in loop quantum cosmology and that in the one-vertex model of quantum-reduced loop gravity, the Lorentzian part of the latter operator being derived from the scalar curvature operator introduced in \cite{Lewandowski:2021iun, Lewandowski:2022xox}. As such, the new proposal should rather be seen as a heuristic conjecture, and not yet as the outcome of any systematic construction or derivation resting on the framework of full loop quantum gravity. For the Euclidean part of the Hamiltonian, it has been shown that an expression corresponding to the polymerized LQC Hamiltonian can be recovered as the expectation value of the operator in the full theory on semiclassical states describing a homogeneous and isotropic spatial geometry \cite{Dapor:2017gdk, Zhang:2021qul}, but no similar calculation is currently available for the curvature operator of \cite{Lewandowski:2021iun, Lewandowski:2022xox}. However, it seems plausible that one can reliably anticipate the result of such a calculation on the basis of the form taken by the curvature operator in the one-vertex model. At the level of computing expectation values, there is unlikely to be any crucial difference between a cubic graph consisting of a single six-valent vertex (with essentially periodic boundary conditions) and a larger cubic graph carrying a coherent state peaked on identical classical data on each edge.

As a first preliminary look into the physical content of the new model, we examined the effective dynamics of the model and compared it against the dynamics of standard LQC and the Dapor--Liegener model, an earlier proposal of similar nature, where a Lorentzian term for the LQC Hamiltonian is derived from Thiemann's regularization of the Hamiltonian constraint in loop quantum gravity. As in standard LQC, the classical singularity is resolved and replaced with a ``bounce'' under the dynamics of the new model, while the trajectory agrees with the classical dynamics of a homogeneous and isotropic universe in the far future and far past of the bounce. The solution derived from the new model is time-symmetric with respect to the past and future of the bounce, in contrast to the Dapor--Liegener model, where a trajectory converging to the classical solution in the far future instead features a de Sitter-like phase of exponential contraction in the past of the bounce. The specific details of the bounce are nevertheless quite different under the new model in comparison with the standard LQC scenario. Most notably, the bounce occurs at a significantly lower value of the volume. It also seems that the duration of the non-classical phase, which interpolates between the classical solutions in the far past and far future, is very slightly shorter in the new model.

At this stage several questions still remain for future work. These include analyzing the new model in terms of its full quantum dynamics, as opposed to merely the effective dynamics, as well as establishing the physical implications of the model regarding cosmological perturbations and the CMB power spectrum, as done for the Dapor--Liegener model e.g.~in \cite{Agullo:2018wbf, Gomar:2020orw, Garcia-Quismondo:2020wna}, in order to determine whether the physical predictions of the new model contain any potentially observable differences in relation to standard LQC. A question of a different kind has to do with quantization ambiguities in the construction of the scalar curvature operator of \cite{Lewandowski:2021iun, Lewandowski:2022xox} and their potential effect on the form of the one-vertex operator \eqref{eq:H_L}. While standard factor ordering ambiguities seem unlikely to be relevant in the semiclassical setting of effective dynamics and coherent state expectation values, there is also a different type of ambiguity, which consists of using identities like $E^a_i\partial_b E_c^i = -E_c^i\partial_b E^a_i$ to write the classical expression of the Ricci scalar as a function of the Ashtekar variables in different forms that are classically equal to each other but lead to different quantum operators upon quantization. By systematically classifying these ambiguities and the resulting curvature operators on the one-vertex graph, one could attempt to establish the full range of polymerized expressions, of which Eq.~\eqref{eq:H_L^mu} is just one specific example, which can be seen as representing the quantization of the scalar curvature in loop quantum cosmology.

\subsection*{Acknowledgments}

This work was funded by National Science Centre, Poland through grant no.~2022\slash 44\slash C\slash ST2\slash 00023.

\printbibliography

\end{document}